\documentclass[aps,prl, twocolumn,superscriptaddress,nofootinbib]{revtex4}
\usepackage{amsmath}
\usepackage{graphicx}
\usepackage{amssymb}
\usepackage[colorlinks,citecolor=blue]{hyperref}

\newcommand{\beq}{\begin{equation}}
\newcommand{\eeq}{\end{equation}}
\newcommand{\bea}{\begin{eqnarray}}
\newcommand{\eea}{\end{eqnarray}}

\newcommand{\nn}{\nonumber}

\allowdisplaybreaks[1]

\begin{document}

\title{A Boost Test of Anomalous Diphoton Resonance at the LHC}

\author{Qing-Hong Cao}
\email{qinghongcao@pku.edu.cn}
\affiliation{Department of Physics and State Key Laboratory of Nuclear Physics and Technology, Peking University, Beijing 100871, China}
\affiliation{Collaborative Innovation Center of Quantum Matter, Beijing 100871, China}
\affiliation{Center for High Energy Physics, Peking University, Beijing 100871, China}

\author{Yandong Liu}
\email{ydliu@pku.edu.cn}
\affiliation{Department of Physics and State Key Laboratory of Nuclear Physics and Technology, Peking University, Beijing 100871, China}

\author{Ke-Pan Xie}
\email{kpxie@pku.edu.cn}
\affiliation{Department of Physics and State Key Laboratory of Nuclear Physics and Technology, Peking University, Beijing 100871, China}

\author{Bin Yan}
\email{binyan@pku.edu.cn}
\affiliation{Department of Physics and State Key Laboratory of Nuclear Physics and Technology, Peking University, Beijing 100871, China}

\author{Dong-Ming Zhang}
\email{zhangdongming@pku.edu.cn}
\affiliation{Department of Physics and State Key Laboratory of Nuclear Physics and Technology, Peking University, Beijing 100871, China}

\begin{abstract}
The recently observed diphoton resonance around 750~GeV at the LHC Run-2 could be interpreted as a weak singlet scalar. The scalar might also decay into a pair of $Z$-boson and photon. The $Z$-boson is highly boosted and appears as a fat jet in the detector. We use the jet substructure method to explore the possibility of discovering the singlet scalar in the process of $pp\to S\to Z\gamma$ in the future LHC experiment.
\end{abstract}

\maketitle

\noindent {\bf Introduction:~}%
Recently, an anomalous resonance around $750~{\rm GeV}$ is observed in the diphoton channel at the level of $3.9\sigma$ by the ATLAS collaboration and $2.6\sigma$ by the CMS collaboration at the LHC Run-II~\cite{atlas:2015temp,CMS:2015dxe}. The observation stimulates great interests in the field~\cite{Harigaya:2015ezk,Mambrini:2015wyu,Backovic:2015fnp,Angelescu:2015uiz,Nakai:2015ptz,Knapen:2015dap,
Buttazzo:2015txu,Pilaftsis:2015ycr,Franceschini:2015kwy,DiChiara:2015vdm,Higaki:2015jag,McDermott:2015sck,Ellis:2015oso,
Low:2015qep,Bellazzini:2015nxw,Gupta:2015zzs,Petersson:2015mkr,Molinaro:2015cwg,Bai:2015nbs,Aloni:2015mxa,
Falkowski:2015swt,Csaki:2015vek,Chakrabortty:2015hff,Bian:2015kjt,Curtin:2015jcv,Fichet:2015vvy,Chao:2015ttq,Demidov:2015zqn,
No:2015bsn,Becirevic:2015fmu,Martinez:2015kmn,Ahmed:2015uqt,Cox:2015ckc,Kobakhidze:2015ldh,Matsuzaki:2015che}. It is attractive to interpret the diphoton resonance as a weak singlet scalar ($S$) which democratically couples to gauge bosons in the Standard Model (SM) through a set of effective operators. There are only three such operators that couple the scalar $S$ to pairs of vector bosons at the dimension five~\cite{Cao:2009ah,Low:2012rj}:
\bea
\mathcal{L}_{eff}&=&\kappa_g \frac{S}{M_S}G^a_{\mu\nu}G^{a\mu\nu} + \kappa_W\frac{S}{M_S}W^i_{\mu\nu}W^{i\mu\nu} \nn\\
&+& \kappa_B\frac{S}{M_S}B_{\mu\nu}B^{\mu\nu},
\eea
where $G_{\mu\nu}^a$, $W_{\mu\nu}^i$ and $B_{\mu\nu}$ denotes the field strength tensor of the $SU(3)_C$, $SU(2)_{W}$ and $U(1)_Y$ gauge group, respectively. Note that the coefficients $\kappa_{g,W,B}$ are expected to be around $\mathcal{O}(M_S/\Lambda)$ with $\Lambda$ being new physics scale beyond the capability of current LHC. Our study can be extended to the weak singlet pesudo-scalar whose couplings to gauge bosons arise from a Wess-Zumino-Witten anomaly term~\cite{Cacciapaglia:2015nga}. After symmetry breaking the SM gauge bosons are interwoven such that the $S$ scalar will also decay into pairs of $WW$, $ZZ$, and $Z\gamma$. 
Observing a resonance in the invariant mass spectrum of $WW$, $ZZ$ and $Z\gamma$ pairs would consolidate the diphoton anomaly. The hadronic modes of the $W$ and $Z$ boson decay are preferred as they exhibit large branching ratios.  On the other hand, the $W$ or $Z$ boson from the $S$ decay is highly boosted such that the two partons from $W$ and $Z$ decays tend to be collimated and appear in the detector as one fat jet, named as a $V$-jet where $V=W/Z$. In this Letter we utilize the so-called jet substructure method to probe the signature of boosted $V$-jets from the $S$ decay in the process of $pp\to S \to Z\gamma$ to crosscheck the diphoton excess.  

\noindent{\bf Scalar production and decay:~}%
We adapt narrow width approximation (NWA) to parameterize the process of $pp\to S \to XY$ as following
\beq
\sigma(pp\to S\to XY) = \sigma(pp\to S)\times\frac{\Gamma(S\to XY )}{\Gamma_{S}},
\eeq
where $X$ and $Y$ denote the SM gauge bosons while $\Gamma_S$ the total width of the $S$ scalar.
The scalar can decay into five modes induced by the three effective operators. The partial widths of the $S$ decay are listed as follows:
\bea
\Gamma(S\to g g)&=&\dfrac{M_S}{\pi}2\kappa_g^2, \nn \\
\Gamma(S\to \gamma\gamma)&=&\dfrac{M_S}{4\pi}\left(\kappa_W s_W^2+\kappa_B c_W^2\right)^2,\nn\\
\Gamma(S\to Z\gamma)&=&\dfrac{M_S}{2\pi}\left(\kappa_W-\kappa_B\right)^2\left(1-r_Z\right)^3 c_W^2s_W^2,\nn\\
\Gamma(S\to WW)&\simeq &\dfrac{M_S}{2\pi}\left(1-6r_W\right)\kappa_W^2,\nn\\
\Gamma(S\to ZZ)&\simeq &\dfrac{M_S}{4\pi}\left(\kappa_Wc_W^2+\kappa_Bs_W^2\right)^2\left(1-6r_Z\right),\nn
\eea
where $r_{V}=m_V^2/M_S^2$.
For a 750~GeV scalar, $r_V(\sim 0.01)$ can be ignored in the above partial widths. We compare the branching ratio $\Gamma(S\to XY)$ to $\Gamma(S\to\gamma\gamma)$ in the following four special cases: \\
i) $\kappa_B=0$, 
\bea
R_{WW}&\equiv &\dfrac{\Gamma(S\to WW)}{\Gamma(S\to\gamma\gamma)}\sim 40, \nn\\
R_{ZZ}&\equiv &\dfrac{\Gamma(S\to ZZ)}{\Gamma(S\to\gamma\gamma)}\sim 12,\nn\\
R_{Z\gamma} &\equiv &\dfrac{\Gamma(S\to Z\gamma)}{\Gamma(S\to\gamma\gamma)}\sim 7;
\eea
ii) $\kappa_W=0$, 
\beq
R_{WW}=0, \quad R_{ZZ}\sim 0.09~, \quad R_{Z\gamma}\sim 0.6~;
\eeq
iii) $\kappa_W=\kappa_B$, 
\bea
R_{WW}=2R_{ZZ} \sim 2, \qquad R_{Z\gamma}=0;
\eea
iv) $\kappa_W=-\kappa_B$,
\beq
R_{WW}\sim 6.9~, \quad R_{ZZ}\sim 1,\qquad R_{Z\gamma} \sim 4.9~.
\eeq
Large $R_{WW}$, $R_{ZZ}$ and $R_{Z\gamma}$ are needed for a discovery of the $S$ scalar in the processes of $pp\to S\to WW/ZZ/Z\gamma$ at the LHC. However, the parameter spaces of $\kappa_{g,W,B}$ are constrained severely by the LHC Run-I data, e.g. $\sigma(WW)\leq 40~{\rm fb}$~\cite{Aad:2015agg}, $\sigma(ZZ)\leq 12~{\rm fb}$~\cite{Aad:2015kna} and $\sigma(Z\gamma)\leq 4~{\rm fb}$~\cite{Aad:2014fha}. After fixing $\sigma(pp\to S\to\gamma\gamma) = 10~{\rm fb}$ to explain the diphoton anomaly~\cite{atlas:2015temp,CMS:2015dxe},  one can convert the cross section bounds into constraints on the ratio $R_{XY}$ as~\cite{Belyaev:2015hgo} 
\beq
R_{WW}\leq 19,\quad R_{ZZ}\leq 6,\quad R_{Z\gamma}\leq 2.
\eeq
Obviously, the two cases of $\kappa_B=0$ and $\kappa_W=-\kappa_B$ have a tension with the current experimental data. It is worth mentioning that the $R_{XY}$ limits are no longer valid if the diphoton excess turns out to be a statistical fluctuation. If the diphoton excess is indeed confirmed in the future experiments, then those two simple cases are excluded. In the following $R_{XY}$ should be treated as the cross section $\sigma(pp\to S\to XY) = R_{XY}\times 10~{\rm fb}$. 

~\\
\noindent{\bf Collider simulation:~}%
Now we turn to collider simulation. The hadronic decay of $WW$ and $ZZ$ from 750 GeV $S$ decay are overwhelmed by the QCD backgrounds and difficult to be trigged at the LHC.  The hard photon in the $Z\gamma$ mode offers a good trigger of the signal events. We thus focus on the $Z(\to jj)\gamma$ mode hereafter. 

For illustration we choose $\kappa_g=\kappa_W=-\kappa_B=0.01$ as our benchmark parameters, which yield the reference cross section and the total width of the $S$ scalar as follows:
\beq
\sigma_0(Z\gamma)= 42.07~{\rm fb},\quad \Gamma^0_{S}=0.07~{\rm GeV}. 
\eeq
Taking advantage of the NWA, the cross section of $pp\to S\to Z\gamma$ for other parameters can be obtained from 
\beq
\sigma(pp\to S\to Z\gamma)=\sigma_0(Z\gamma)\left(\dfrac{\kappa_g}{0.01}\right)^2 \left(\dfrac{\kappa_W-\kappa_B}{0.02}\right)^2 \dfrac{\Gamma^0_{S}}{\Gamma_{S}}.
\label{eq:master}
\eeq
Even though our study is based on the NWA, the results are valid for a large-width scalar, e.g. $\Gamma_S = 0.06 M_S \sim 45~{\rm GeV}$.

The $Z$ boson in the scattering of $pp\to S\to Z\gamma$ tends to be highly boosted. The distance of two partons from the $Z$-boson decay can be estimated approximately as 
\beq
\Delta R \sim 2M_Z/p_T \sim 4M_Z/M_S \sim 0.4-0.5~~,
\eeq
where $\Delta R_{ij} = \sqrt{(\eta_i-\eta_j)^2 + (\phi_i-\phi_j)^2}$ with $\eta_i$ and $\phi_i$ denoting the rapidity and azimuthal angle of parton $i$. Given such a small angular separation, the hadronic decay products from the $Z$-boson would form a fat jet with a substructure in the detector. That yields a special collider signature of one fat $Z$-jet and one hard photon.
In order to mimic the signal events, the SM background should consist of $W$ or $Z$ boson and a hard photon. We consider SM backgrounds as follows: i) the associated production of a $W$ boson and a photon (denoted by $W$+$\gamma$); ii) the associated production of a $Z$ boson and a photon ($Z$+$\gamma$); iii) the associated production of a photon and multiple jets (denoted by $\gamma$+jets). The other backgrounds such as $W$+jets, $Z$+jets and $t\bar{t}$ are highly suppressed by demanding a hard photon in the final state. 

We generate both the signal and the background processes at the parton level using MadEvent~\cite{Alwall:2007st} at the 14~TeV LHC and pass events to Pythia~\cite{Sjostrand:2014zea} for showering and hadronization. The Delphes package~\cite{deFavereau:2013fsa} is used to simulate detector smearing effects in accord to a fairly standard Gaussian-type detector resolution given by $\delta E/E= \mathcal{A}/\sqrt{E/{\rm GeV}}\oplus \mathcal{B}$,
where $\mathcal{A}$ is a sampling term and $\mathcal{B}$ is a constant term.  For leptons we take $\mathcal{A}=5\%$ and $\mathcal{B}=0.55\%$, and for jets we take $\mathcal{A}=100\%$ 
and $\mathcal{B}=5\%$.
We also impose the lepton veto if the lepton has transverse momentum ($p_T$) greater than $20$ GeV, rapidity $\left|\eta_{\ell}\right|\leq 2.5~$ and its overlap with jets  $\Delta R_{j\ell} \geq 0.4$. 

\begin{figure}
\includegraphics[scale=0.2]{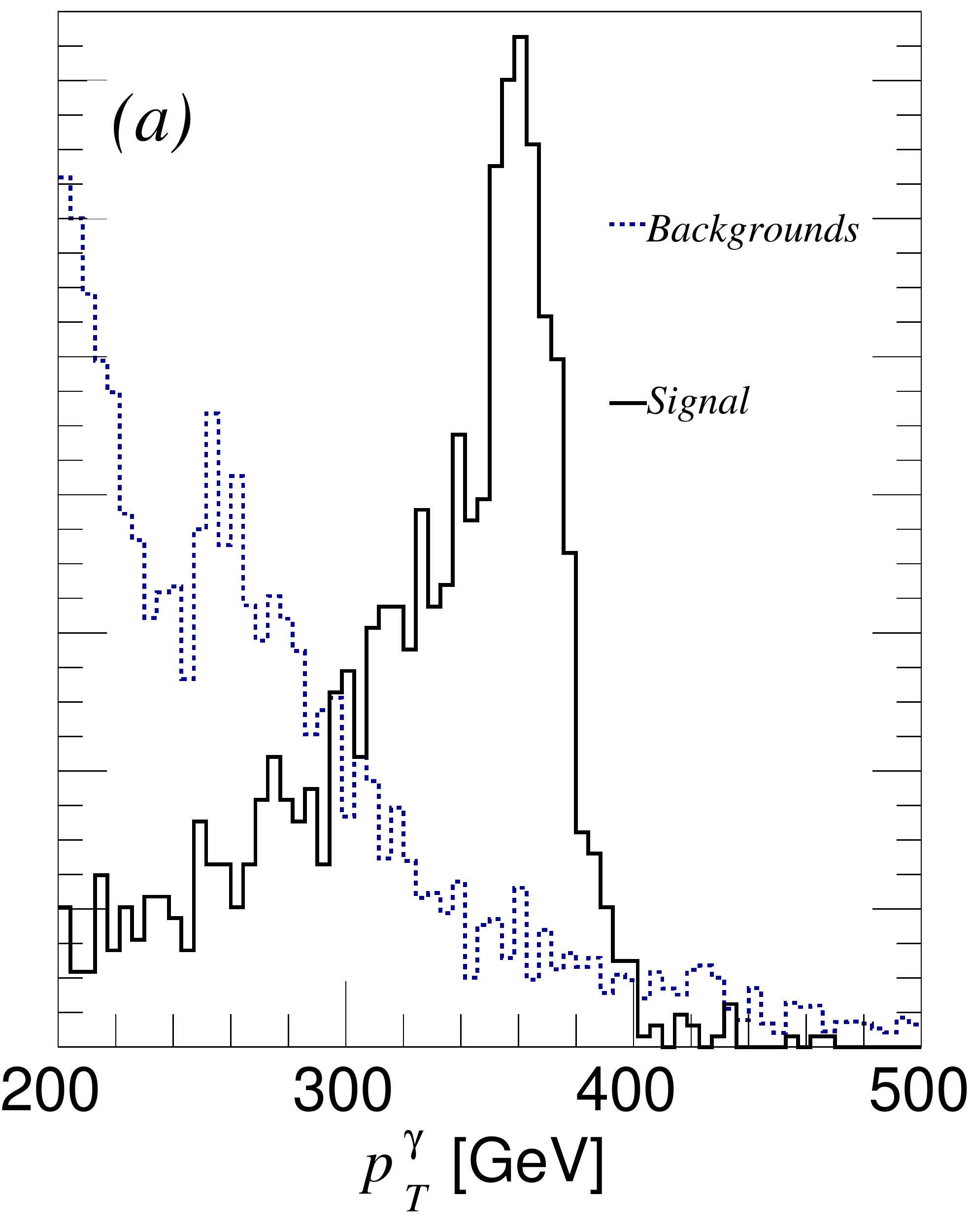}
\includegraphics[scale=0.2]{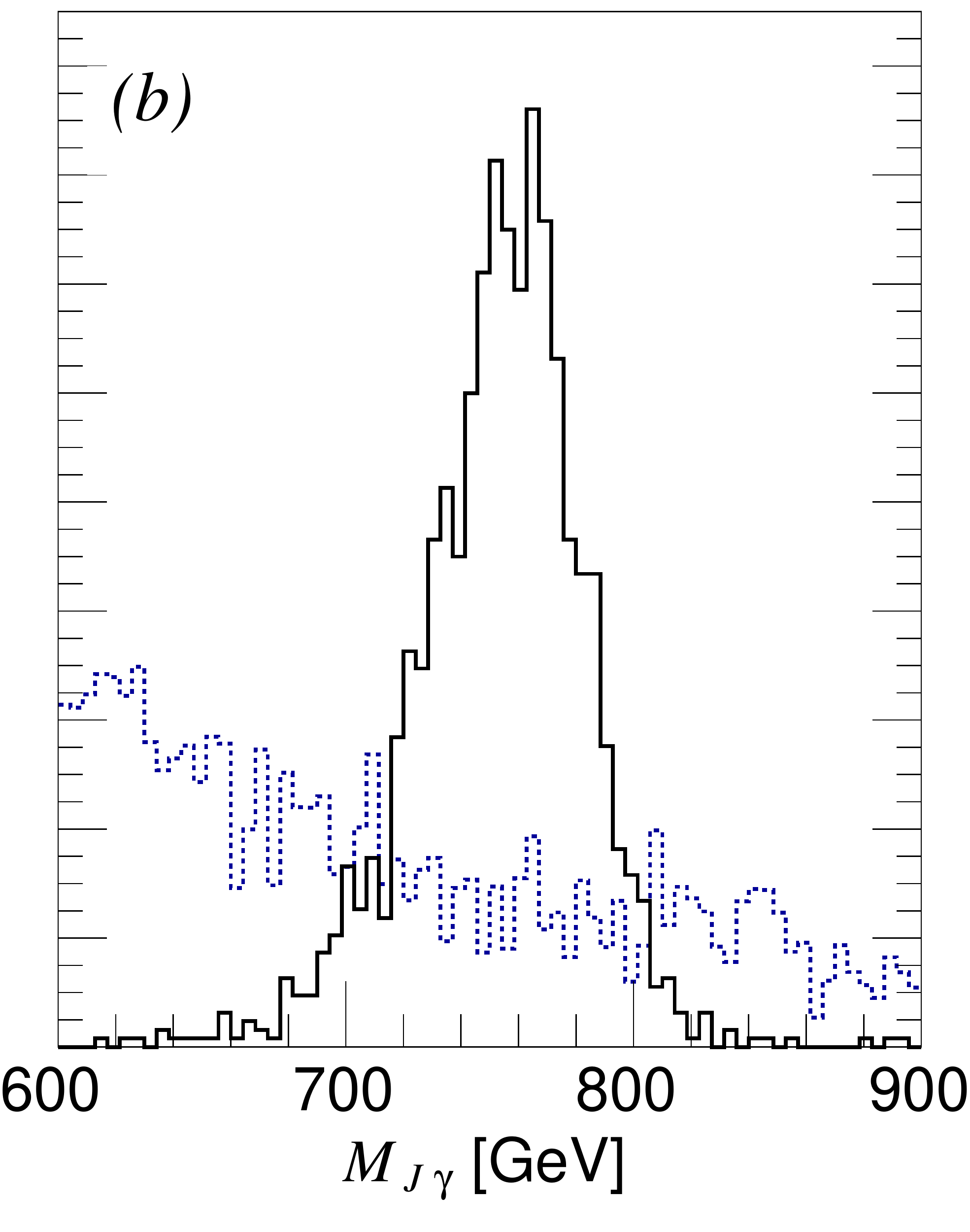}
\caption{The normalized $p_T$ distributions of the photon (a) and  the invariant mass distribution of the $Z$-jet and the photon pair (b).}
\label{fig:dist}
\end{figure}

In the signal event, the photon arises from the heavy scalar decay and thus exhibits a hard peak in the transverse momentum distribution. As sharing the energy with the associated $Z$-boson, the $p_T$ distribution of the photon peaks around $\sim m_S/2 \approx 375$ GeV; see Fig.~\ref{fig:dist}(a). In the analysis we require tagging a hard photon in the final state which satisfies
\beq
p_T^\gamma \ge 250 ~{\rm GeV}, ~~|\eta^\gamma| \le 1.4~~.
\eeq
A 2-pronged boosted $Z$-jet is tagged using the so-called ``mass-drop" technique with asymmetry cut introduced in Ref.~\cite{Butterworth:2008iy}. The $Z$-jet reconstruction is performed using Cambridge/Aachen algorithm with Fastjet~\cite{Cacciari:2011ma}. The distance parameter of $1.2$ is used to cluster a fat jet that is initiated by the boosted $Z$-boson.  
We further require the invariant mass of the reconstructed $Z$-jet ($M_J$) within mass window~\cite{Aad:2015owa}:
\beq
\left| M_{J} -m_Z\right| \leq 13~{\rm GeV}
\eeq
where $m_Z=91.2~{\rm GeV}$. 
Furthermore, the invariant mass of the reconstructed $Z$-jet and the photon is required to lie within the mass window 
\beq
\Delta M_{J\gamma}\equiv \left| M_{J \gamma} - M_S\right|\leq 25~{\rm GeV}.
\eeq
Figure~\ref{fig:dist}(b) plots the invariant mass distribution of the reconstructed $Z$-jet and $\gamma$.
The numbers of events of the signal for our benchmark parameter and the backgrounds after all the above cuts are shown in the fourth  row of Table~\ref{tab:events} with an integrated luminosity ($\mathcal{L}$) of $1~{\rm fb}^{-1}$. Cross sections of other parameters can be derived from Eq.~\ref{eq:master} and those numbers given in Table~\ref{tab:events}.  The cut efficiency of the signal event is not sensitive to the values of $\kappa_{g,W,B}$ or the narrow width of the $S$ scalar. After all the cuts the major background is from the productions of $\gamma$+jets.

\begin{table}
\caption{The numbers of the signal ( $\kappa_{g}=\kappa_{W}=-\kappa_{B}=0.01$ ) and background events  after kinematic cuts at the 14~TeV LHC with an integrated luminosity of 1 fb$^{-1}$.}
\label{tab:events}
\begin{tabular}{c|c|c|c|c} 
\hline
 &  Signal &  $\gamma$+jets & $Z$+$\gamma$ & $W$+$\gamma$ \\ \hline
 No cut    &		42.07			&	454200  & 		2255.5	 &4690\\ \hline
 $p_T^\gamma$ cut and $Z$-jet tagging  &  	1.15	& 179.46 	& 6.15    	 &3.63\\ \hline
 $M_{J\gamma }$ cut& 	0.72	&10.85		& 0.43		 &0.27\\ \hline
\end{tabular}
\end{table}

\begin{figure}
\includegraphics[scale=0.24]{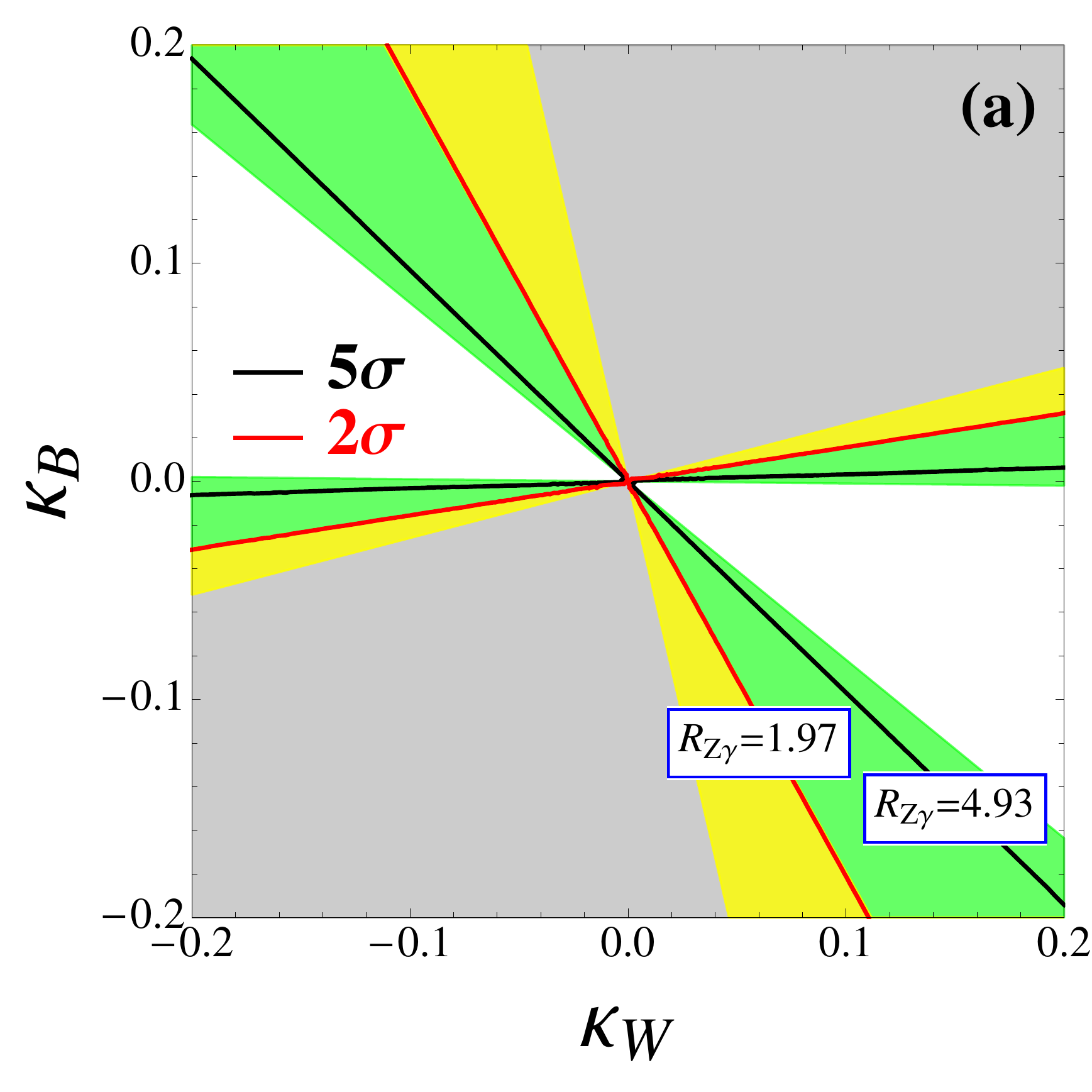}
\includegraphics[scale=0.24]{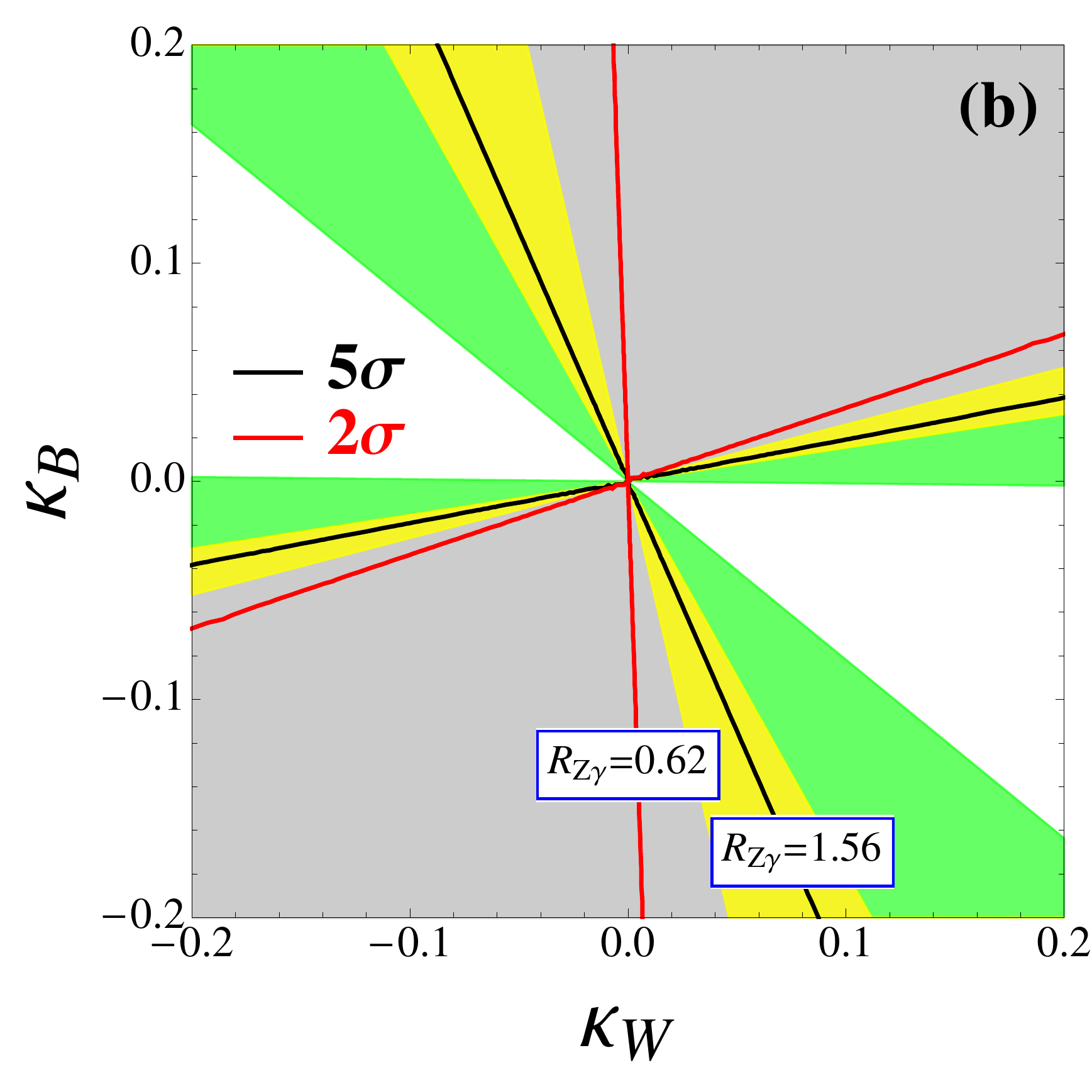}
\caption{The discovery potential of the process of $pp \to S\to Z\gamma$ at the 14~TeV LHC with the integrated luminosity of $300~{\rm fb}^{-1}$ (a) and  $3000~{\rm fb}^{-1}$ (b). The black curves label the $5\sigma$ discovery potential while the red curves represent the $2\sigma$ exclusion. The gray (yellow, green) region denotes the parameter space of $R_{Z\gamma}<1$ ($1<R_{Z\gamma}<2$, $2<R_{Z\gamma}<7$), respectively. }
\label{fig:lhc}
\end{figure}

As the numbers of the signal and background events are large, we estimate the needed cross section of the signal to claim a $5\sigma$ discovery from 
\beq
\sigma_{Z\gamma}=\frac{5\times \sqrt{\sigma_B}}{\epsilon_{\rm cut}\times \sqrt{\mathcal{L}}}.
\eeq
where $\epsilon_{\rm cut}\sim 0.02$ denotes the cut efficiency of the signal. Figure~\ref{fig:lhc}(a) displays the discovery potential of the 750~GeV scalar in the process of $pp\to S\to Z\gamma$  at the 14~TeV LHC with an integrated luminosity of $300~{\rm fb}^{-1}$. The gray (yellow, green) region represents $R_{Z\gamma}\leq 1$ ($1<R_{Z\gamma}<2$, $2<R_{Z\gamma}<7$), respectively.  A large ratio $R_{Z\gamma}=4.93$ is needed to reach a discovery at the level of $5\sigma$; see the black curve. If no excess were observed, then one can exclude the parameter spaces of $R_{Z\gamma}>1.97$ at the $2\sigma$ level; see the red curve. 
The high luminosity phase of the LHC ($\mathcal{L}=3000~{\rm fb}^{-1}$) could probe the parameter spaces of $R_{Z\gamma}\geq 1.56$ at the $5\sigma$ level and exclude the parameter spaces of $R_{Z\gamma}\geq 0.62$ at the $2\sigma$ level; see Fig.~\ref{fig:lhc}(b).

~\\
\noindent{\bf Acknowledgement:~}The work is supported in part by the National Science Foundation of China under Grand No. 11275009.

\bibliographystyle{apsrev}
\bibliography{reference}

\end{document}